\documentclass[lettersize,journal]{IEEEtran}
\usepackage{amsmath,amsfonts}
\usepackage{algorithmic}
\usepackage{algorithm}
\usepackage{array}
\usepackage[caption=false,font=footnotesize,labelfont=rm,textfont=rm]{subfig}
\usepackage{textcomp}
\usepackage{stfloats}
\usepackage{url}
\usepackage{verbatim}
\usepackage{graphicx}
\usepackage{cite}
\usepackage{stfloats}
\usepackage{makecell}
\usepackage{multirow}
\usepackage{amsthm,amsmath,amssymb}
\usepackage{mathrsfs}
\usepackage{epsfig}
\usepackage{epstopdf}
\usepackage{diagbox}
\usepackage{soul}
\usepackage{color, xcolor}
\usepackage{pifont}
\begin{document}

\title{How to Extend 3D GBSM to Integrated Sensing and Communication Channel with Sharing Feature?}
%

\author{Yameng Liu, Jianhua Zhang, Yuxiang Zhang, Huiwen Gong, Tao Jiang, Guangyi Liu
}

\markboth{Journal of \LaTeX\ Class Files,~Vol.~14, No.~8, August~2021}%
{Shell \MakeLowercase{\textit{et al.}}: A Sample Article Using IEEEtran.cls for IEEE Journals}


\maketitle

\begin{abstract}
Integrated Sensing and Communication (ISAC) is a promising technology in 6G systems. The existing 3D Geometry-Based Stochastic Model (GBSM), as standardized for 5G systems, addresses solely communication channels and lacks consideration of the integration with sensing channel. Therefore, this letter extends 3D GBSM to support ISAC research, with a particular focus on capturing the sharing feature of both channels, including shared scatterers, clusters, paths, and similar propagation parameters, which have been experimentally verified in the literature. The proposed approach can be summarized as follows: Firstly, an ISAC channel model is proposed, where shared and non-shared components are superimposed for both communication and sensing. Secondly, sensing channel is characterized as a cascade of TX-target, radar cross section, and target-RX, with the introduction of a novel parameter $\bf{S}$ for shared target extraction. Finally, an ISAC channel implementation framework is proposed, allowing flexible configuration of sharing feature and the joint generation of communication and sensing channels. The proposed ISAC channel model can be compatible with the 3GPP standards and offers promising support for ISAC technology evaluation. 

\end{abstract}


\section{Introduction}\label{section1}
\IEEEPARstart{T}{he} sixth generation (6G) systems envision intelligently connecting everything, which requires a high transmission rate and precise sensing accuracy \cite{zhang2020channel}. Integrated sensing and communication (ISAC) has been identified as a typical usage scenario for 6G \cite{itu2023framework}. Different from conventional systems with separate functions, ISAC technology allows the base stations or terminals to communicate with each other and sense their surroundings. By sharing the same frequency and hardware resources, ISAC systems offer advantages in improving spectrum utilization and reducing costs \cite{liu2022shared,ouyang2022performance}.

Realistic channel modeling is the prerequisite for the design of every generation of systems. The mainstream channel model, Geometry-Based Stochastic channel Model (GBSM), is the standardized modeling method adopted by ITU and 3GPP. It is two-dimensions (2D) in 4G systems and introduced the vertical angles and expanded to 3D in 5G systems \cite{zhang20173}. GBSM is a clustered structure model, grouping multipaths into stochastic clusters. However, this channel model focuses on communication channel, lacking considering the integration with sensing channel. On the one hand, stochastic clusters in 3D GBSM cannot accurately represent the identified locations and attributes of the targets, which are crucial for sensing applications such as positioning and environment reconstruction. On the other hand, due to the multiplexing of hardware resources and the same propagation environment, some targets might serve as shared scatterers for both the communication and sensing channels. This sharing feature, including shared scatterers, clusters, paths, and similar propagation parameters, should be included in ISAC channel model to evaluate the integrated functions and algorithm design of ISAC systems.

With the development of the integrated systems, some existing ISAC studies have operated under the assumption of the sharing feature/correlation between communication and sensing channels \cite{liu2020joint,liu2022survey,fan2022radar}. In \cite{ali2020passive,gonzalez2016radar}, the similarity of the azimuth power spectrum of communication and sensing are investigated using Ray-tracing simulations. To validate the existence of the sharing feature, we conduct indoor ISAC channel measurements at 28 GHz in \cite{liu2022shared}, intuitively observing the shared scatterers from the power-angular-delay profiles of communication and sensing channels. 
To further extend GBSM to ISAC channels, the work in \cite{lopez2022considering} analyzes the correlation between communication and sensing channels under realization of GBSM. In \cite{liu2022shared}, we propose a shared cluster-based stochastic ISAC channel model according to the measured observation. However, a practical ISAC channel model that extends 3D GBSM with consideration for the sharing feature is still missing in the literature.

To bridge these gaps, this letter proposes a 3D GBSM principle based ISAC channel model with sharing feature. In this model, communication and sensing Channel Impulse Responses (CIRs) are composed of shared and non-shared components. A novel parameter $\bf{S}$ is introduced for shared target/cluster extraction and parametric modeling of the sensing channel is presented. Then, we propose an ISAC channel implementation framework, which is compatible with the 3GPP standards. The sensing cluster/path positions, the relationship between shared parameters, and the statistical sharing degree (SD) are analyzed to validate the model performance.

\section{ISAC Channel Modeling}\label{section2}

\begin{figure}[!h]
\centering
\includegraphics[width=3.4in]{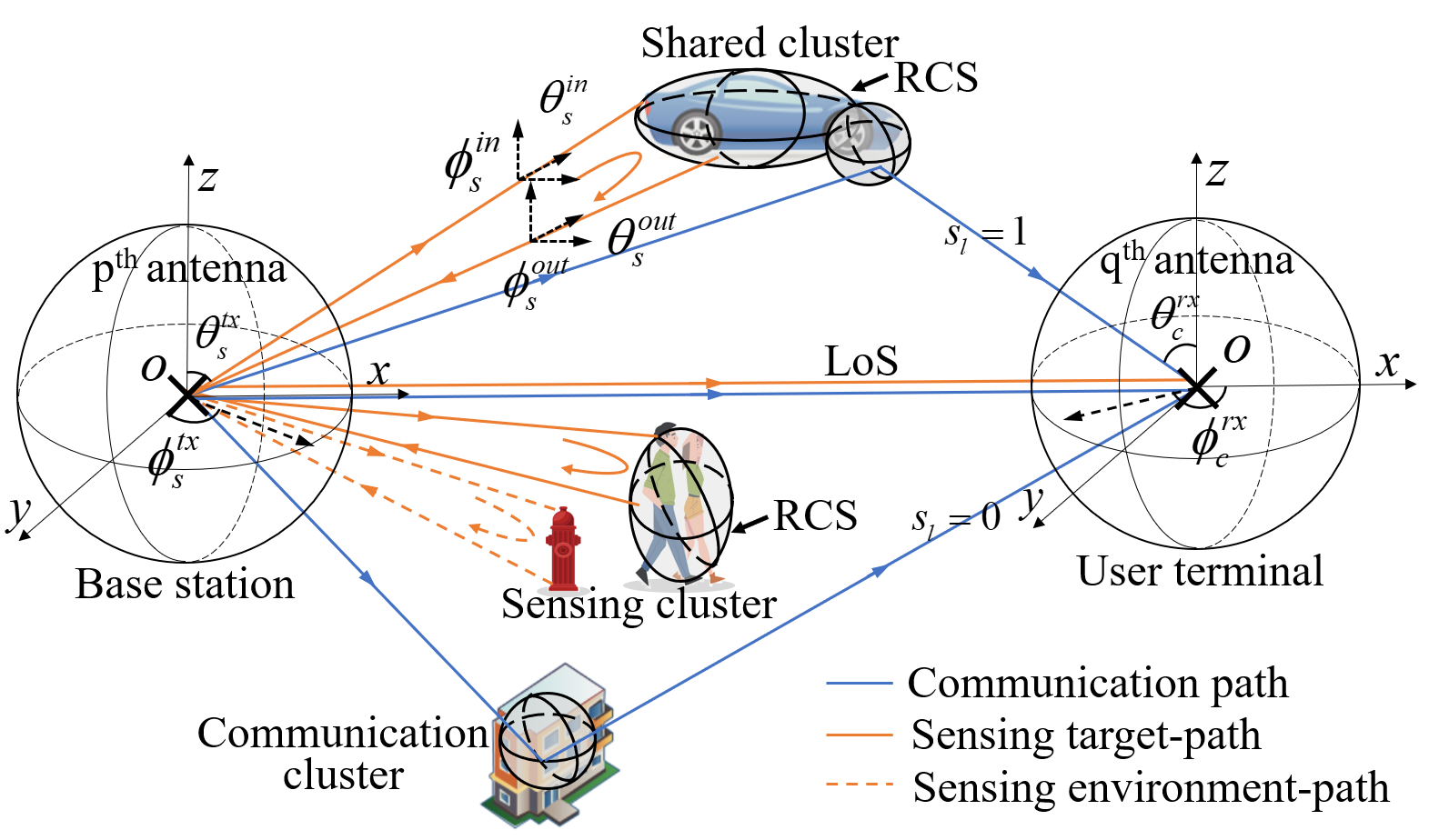}
\caption{The illustration of ISAC channel model. The blue lines denote communication paths, and the red solid and dashed lines respectively denote sensing target- and environment- paths.}
\label{fig_1}
\end{figure}

This letter considers a wideband ISAC channel under a frequency selective fading channel. Fig. \ref{fig_1} provides an illustration of the proposed ISAC channel model. The Base Station (BS) and User Terminal (UT) are respectively equipped with $P$ and $Q$ antennas, serving as the transmitter (TX) and receiver (RX) for communication. The sensing TX and RX are equipped with $E$ and $W$ antennas. As shown in Fig. \ref{fig_1}, assuming that BS serves as sensing TX ($E=P$), mono-static sensing means the sensing signals are received as the echo, i.e., $W=P$. When the UT or other terminal serve as the sensing RX, i.e., $W=Q\ \rm{or}\ \rm{else}$, it represents bi-static sensing mode. Communication and sensing signals will contribute some shared clusters, also non-shared communication or sensing clusters.




\subsection{ISAC channel model} 

In the proposed model, We designate the sensing targets of interest are indexed as $l=1,2,...,L$. The cluster number as $n=1,2,...,N$, and ray number within each cluster as $m=1,2,...,M$. To capture the sharing feature in ISAC channel model, we express the communication and sensing CIRs as the superposition of shared and non-shared components. The CIRs at time $t$ between TX antenna $p/e$ and RX antenna $q/w$ for communication/sensing are respectively formulated as

\begin{equation}
\label{eqn_1}
{H_{c,q,p}(t,\tau_c)}=H_{c,q,p}^{{\rm{sc}}}(t,\tau_c)+H_{c,q,p}^{{\rm{nsc}}}(t,\tau_c),
\end{equation}
\begin{align}
\label{eqn_2}
H_{s,w,e}(t,\tau_s)=&\sum\limits_{{l}}^{{L}} {{H_{s,w,l,e}}(t,{\tau _s})} \notag\\
=&H_{s,w,e}^{{\rm{sc}}}(t,\tau_s)+H_{s,w,e}^{{\rm{nsc}}}(t,\tau_s),
\end{align}
where ${H_{s,w,l,e}}(t,\tau_s)$ denotes the propagation from TX$_c$ experience the $l^{th}$ target to RX$_c$. $H_{c,q,p}^{{\rm{sc}}}(t,\tau_c)$ and $H_{c,q,p}^{{\rm{nsc}}}(t,\tau_c)$ denote the shared and non-shared components in communication channel. $H_{s,w,e}^{{\rm{sc}}}(t,\tau_s)$ and $H_{s,w,e}^{{\rm{nsc}}}(t,\tau_s)$ denote the corresponding components in sensing channel. The expanded expressions have been given as (\ref{eqn_3}). In these formulas, 
\begin{itemize}
\item{$(\cdot)_c$ and $(\cdot)_s$ denote the communication and sensing channels, respectively.}
\item{$(\cdot)^T$ stands for matrix transposition.}
\item{$\lambda_0$ is the wavelength of the carrier frequency.}
\item{$a_{c,{n},{m}}$, ${a_{s,l,{n},{m}}^{tar,w}}$, and ${a_{s,l,{n},{m}}^{e,tar}}$ are the amplitudes of cluster $n^{th}$ path $m^{th}$ in communication channel, target$_l\text{-RX}_s$ and $\text{TX}_s\text{-}$target$_l$ sub-channels.}
\item{$\theta _{c,{n},{m}}^{\rm{zoa}}$, $\phi _{c,{n},{m}}^{\rm{aoa}}$, $\theta _{c,{n},{m}}^{\rm{zod}}$, and $\phi _{c,{n},{m}}^{\rm{aod}}$ represent the Zenith Angle of Arrival (ZAoA), the Azimuth Angle of Arrival (AAoA), the ZAoD, and the AAoD of the cluster $n^{th}$ path $m^{th}$ at communication RX and TX, respectively. $\theta _{s,l,{n},{m}}^{\rm{zoa}}$, $\phi _{s,l,{n},{m}}^{\rm{aoa}}$, $\theta _{s,l,{n},{m}}^{\rm{zod}}$, and $\phi _{s,l,{n},{m}}^{\rm{aod}}$ represent the ZAoA, AAoA, ZAoD, and AAoD of the target $l^{th}$ cluster $n^{th}$ path $m^{th}$ at sensing RX and TX, respectively.}
\item{$\theta _{s,l,{n},{m}}^{{\rm{out}}}$ and $\phi _{s,l,{n},{m}}^{{\rm{out}}}$ are the zenith and azimuth angle of the cluster $n^{th}$ path $m^{th}$ outgoing the sensing target in the target$_l\text{-RX}_s$ sub-channel. $\theta _{s,l,{n},{m}}^{{\rm{in}}}$ and $\phi _{s,l,{n},{m}}^{{\rm{in}}}$ are the zenith and azimuth angle of the cluster $n^{th}$ path $m^{th}$ incident the sensing target in the $\text{TX}_s\text{-}$target$_l$ sub-channel.}
\item{$F_{c,rx,q}^\theta$, $F_{c,rx,q}^\phi$, $F_{c,tx,p}^\theta$, and $F_{c,tx,p}^\phi$ indicate the radiation pattern of the communication RX antenna $q$ and TX antenna $p$, in the $\theta$ and $\phi$ polarization. $F_{s,rx,w}^\theta$, $F_{s,rx,w}^\phi$, $F_{s,tx,e}^\theta$, and $F_{s,tx,e}^\phi$ indicate the radiation pattern of the sensing RX antenna $w$ and TX antenna $e$, in the $\theta$ and $\phi$ polarization.}
\item{$\psi_{c,{n},{m}}^{i j }$ and $\psi_{s,l,{n},{m}}^{i j }$, $i,j\in \{\theta, \phi \}$, respectively denote the random phase of the cluster $n^{th}$ path $m^{th}$ that originate in the $i$ direction and arrive in the $j$ direction in communication CIR and sensing CIR of target $l^{th}$.}
\item{$\kappa _{c,{n},{m}}$ and $\kappa _{s,l,{n},{m}}$ represent the cross polarization power ratio (XPR) of the communication and sensing (target $l^{th}$) cluster $n^{th}$ path $m^{th}$ in the linear scale.}
\item{$\sigma ^{ij}$, $i,j\in \{\theta, \phi \}$ denotes the RCS complex coefficient that originate in $i$ direction and arrive in $j$ direction. }
\item{$\widehat r_{c,rx,{n},{m}}$, $\widehat r_{c,tx,{n},{m}}$, $\widehat r_{s,rx,l,{n},{m}}$, and $\widehat r_{s,tx,l,{n},{m}}$ are the spherical unit vector of communication and sensing (target $l^{th}$) cluster $n^{th}$ path $m^{th}$ at RX and TX sides.}
\item{${\overline d }_{c,q}$, ${\overline d }_{c,p}$, ${\overline d }_{s,w}$, and ${\overline d }_{s,e}$ are the location vectors of antennas $q$ and $p$ at communication sensing RX and TX sides, respectively.}
\item{$f_{d,c,{n},{m}}$ and $f_{d,s,l,{n},{m}}$ denote the Doppler shift of communication and sensing (target $l^{th}$) cluster $n^{th}$ path $m^{th}$.}
\item{$\tau _{c,{n},{m}}$ is the delay of cluster $n^{th}$ path $m^{th}$ in communication channel. $\tau _{s,l,{n},{m}}^{tar,w}$ and $\tau _{s,l,{n},{m}}^{p,tar}$ indicate the delay of cluster $n^{th}$ path $m^{th}$ in sensing target$_l\text{-RX}_s$ and $\text{TX}_s\text{-}$target$_l$ sub-channels.}
\end{itemize}

In (\ref{eqn_3}), the shared target, clusters, and paths are indexed by $l_{c0}$, $n_{c0}/n_{s0}$ and $m_{c0}/m_{s0}$, and the non-shared components of communication and sensing CIRs are represented as cluster $n_1$ path $m_1$ and target $l_2$ cluster $n_2$ path $m_2$, respectively. The total number of communication and sensing clusters can be given as $N_c=N_{c0}+N_1$ and $N_s=L_{s0}\times N_{s0}+L_2\times N_2$.

\begin{figure*}
\begin{small}
\begin{subequations}
\label{eqn_3}
\begin{align}
\label{eqn_3a}
H_{c,q,p}^{{\rm{sc}}}/H_{c,q,p}^{{\rm{nsc}}}(t,\tau_c)=&\sum\limits_{{n_{c0}/n_1}}^{{N_{c0}/N_1}} {\sum\limits_{{m_{c0}/m_1}}^{{M_{c0}/M_1}} {{a_{c,{n},{m}}} {{{\left[ {
\begin{array}{*{20}{c}}
{{F_{c,rx,q}^\theta}({\theta _{c,{n},{m}}^{\rm{zoa}}},{\phi _{c,{n},{m}}^{\rm{aoa}}})}\\
{{F_{c,rx,q}^\phi}({\theta _{c,{n},{m}}^{\rm{zoa}}},{\phi _{c,{n},{m}}^{\rm{aoa}}})}
\end{array}} \right]}^T}\left[ 
\setlength{\arraycolsep}{0.5pt}
{\begin{array}{*{20}{c}}
{\exp (j\psi _{c,{n},{m}}^{\theta \theta })}&{\sqrt {\kappa _{c,{n},{m}}^{ - 1}} \exp (j\psi _{c,{n},{m}}^{\theta \phi })}\\
{\sqrt {\kappa _{c,{n},{m}}^{ - 1}} \exp (j\psi _{c,{n},{m}}^{\phi \theta })}&{\exp (j\psi _{c,{n},{m}}^{\phi \phi })}
\end{array}} \right]} } }\notag\\
&\cdot 
\left[ {\begin{array}{*{20}{c}}
{{F_{c,tx,q}^\theta}({\theta _{c,{n},{m}}^{\rm{zod}}},{\phi _{c,{n},{m}}^{\rm{aod}}})}\\
{{F_{c,tx,q}^\phi}({\theta _{c,{n},{m}}^{\rm{zod}}},{\phi _{c,{n},{m}}^{\rm{aod}}})}
\end{array}} \right]\exp \left( {\frac{{j2\pi (\widehat r_{c,rx,{n},{m}}^{T} \cdot {{\overline d }_{c,q}})}}{{{\lambda _0}}}} \right)\exp \left( {\frac{{j2\pi (\widehat r_{c,tx,{n},{m}}^{T} \cdot {{\overline d }_{c,p}})}}{{{\lambda _0}}}} \right)\notag \\
&\cdot 
\exp \left( {j2\pi {f_{d,c,{n},{m}}}t} \right)\delta ({\tau _c} - {\tau _{c,{n},{m}}}),\\
\label{eqn_3b}
H_{s,w,e}^{{\rm{sc}}}/H_{s,w,e}^{{\rm{nsc}}}(t,\tau_s) 
=&\sum\limits_{{l_{s0}/l_2}}^{{L_{s0}/L_2}} \sum\limits_{{n_{s0}/n_2}}^{{N_{s0}/N_2}} {\sum\limits_{{m_{s0}/m_2}}^{{M_{s0}/M_2}} {{a_{s,l,{n},{m}}^{tar,w}}{a_{s,l,{n},{m}}^{e,tar}}{{\left[ {\begin{array}{*{20}{c}}
{{F_{s,rx,w}^\theta}({\theta _{s,l,{n},{m}}^{\rm{zoa}}},{\phi_{s,l,{n},{m}}^{\rm{aoa}}})}\\
{{F_{s,rx,w}^\phi}({\theta _{s,l,{n},{m}}^{\rm{zoa}}},{\phi_{s,l,{n},{m}}^{\rm{aoa}}})}
\end{array}} \right]}^T}}} \notag\\
&\cdot
{{\left[
\setlength{\arraycolsep}{0.1pt}
{\begin{array}{*{20}{c}}
{\exp (j\psi _{s,l,{n},{m}}^{\theta \theta })}&{\sqrt {\kappa _{s,l,{n},{m}}^{ - 1}} \exp (j\psi _{s,l,{n},{m}}^{\theta \phi })}\\
{\sqrt {\kappa _{s,l,{n},{m}}^{ - 1}} \exp (j\psi _{s,l,{n},{m}}^{\phi \theta })}&{\exp (j\psi _{s,l,{n},{m}}^{\phi \phi })}
\end{array}} \right]} } \notag\\
&\cdot
\left[ {\begin{array}{*{20}{c}}
{{\sigma_l ^{\theta \theta }}(\theta _{s,l,{n},{m}}^{{\rm{out}}},\phi _{s,l,{n},{m}}^{{\rm{out}}},\theta _{s,l,{n},{m}}^{\rm{in}},\phi _{s,l,{n},{m}}^{\rm{in}})}&{{\sigma_l ^{\theta \phi }}(\theta _{s,l,{n},{m}}^{{\rm{out}}},\phi _{s,l,{n},{m}}^{{\rm{out}}},\theta _{s,l,{n},{m}}^{\rm{in}},\phi _{s,l,{n},{m}}^{\rm{in}})}\\
{{\sigma_l ^{\phi \theta }}(\theta _{s,l,{n},{m}}^{{\rm{out}}},\phi _{s,l,{n},{m}}^{{\rm{out}}},\theta _{s,l,{n},{m}}^{\rm{in}},\phi _{s,l,{n},{m}}^{\rm{in}})}&{{\sigma_l ^{\phi \phi }}(\theta _{s,l,{n},{m}}^{{\rm{out}}},\phi _{s,l,{n},{m}}^{{\rm{out}}},\theta _{s,l,{n},{m}}^{\rm{in}},\phi _{s,l,{n},{m}}^{\rm{in}})}
\end{array}} \right] \notag\\
&\cdot 
\left[ {\begin{array}{*{20}{c}}
{{F_{s,tx,p}^\theta}({\theta _{s,l,{n},{m}}^{\rm{zod}}},{\phi _{s,l,{n},{m}}^{\rm{aod}}})}\\
{{F_{s,tx,p}^\phi}({\theta _{s,l,{n},{m}}^{\rm{zod}}},{\phi _{s,l,{n},{m}}^{\rm{aod}}})}
\end{array}} \right]\exp \left( {\frac{{j2\pi (\widehat r_{s,rx,l,{n},{m}}^T \cdot {{\overline d }_{s,w}})}}{{{\lambda _0}}}} \right) \exp \left( {\frac{{j2\pi (\widehat r_{s,tx,l,{n},{m}}^T \cdot {{\overline d }_{s,e}})}}{{{\lambda _0}}}} \right)  \notag\\
&\cdot 
\exp \left( {j2\pi {f_{d,s,l,{n},{m}}}t} \right)\delta ({\tau _s} - \tau _{s,l,{n},{m}}^{{tar,w}} - \tau _{s,l,{n},{m}}^{{e,tar}}).
\end{align}
\end{subequations}
\end{small}
\end{figure*}

\subsection{ISAC Model Parameterization}


This subsection elaborates on how the parameters in the proposed model, which deviate from the traditional 3D GBSM, are depicted. To account for the influence of targets, the sensing channel is modeled by a TX$_s$-target$_l$-RX$_s$ cascade channel, which is composed of the TX$_s$-target$_l$ sub-channel, target$_l$-RX$_s$ sub-channel, and the target response. The calculation can be expressed as
\begin{equation}
\label{eqn_4}
{H_{s,w,l,e}}(t,\tau_s)=H_{s,w,l}(t,\tau_s)* {\sigma}_l * H_{s,l,e}(t,\tau_s),
\end{equation}
where $H_{s,w,l}(t,\tau_s)$ and $H_{s,l,e}(t,\tau_s)$ represent the channel coefficient for target$_l$-RX$_s$ and TX$_s$-target$_l$ sub-channels. ${\sigma}_l$ is a composite RCS model of $l^{th}$ target, expressed as
\begin{equation}
\label{eqn_5}
\sigma_l=\left[\setlength{\arraycolsep}{1.2pt} 
{\begin{array}{*{20}{c}}
{{\sigma_l ^{\theta \theta }}(\theta ^{{\rm{out}}},\phi ^{{\rm{out}}},\theta ^{\rm{in}},\phi ^{\rm{in}})}&{{\sigma_l ^{\theta \phi }}(\theta ^{{\rm{out}}},\phi ^{{\rm{out}}},\theta ^{\rm{in}},\phi ^{\rm{in}})}\\
{{\sigma_l ^{\phi \theta }}(\theta ^{{\rm{out}}},\phi ^{{\rm{out}}},\theta ^{\rm{in}},\phi ^{\rm{in}})}&{{\sigma_l ^{\phi \phi }}(\theta ^{{\rm{out}}},\phi ^{{\rm{out}}},\theta ^{\rm{in}},\phi ^{\rm{in}})}
\end{array}} \right]
\end{equation}
in (\ref{eqn_3b}) to accurately characterize the target response. In this letter, $\sigma_l$ is modeled as a $2\times2$ array with different polarization directions. The $\sigma^{ij}$ values are the function of angles that outgoing ($\theta ^{{\rm{out}}}$ and $\phi ^{{\rm{out}}}$) and incident ($\theta ^{\rm{in}}$ and $\phi ^{\rm{in}}$) the targets.

Due to the cascading nature of the sensing channel, the amplitudes (${a_{s,l,n,m}^{tar,w}}$ and ${a_{s,l,n,m}^{e,tar}}$) and delays ($\tau _{s,l,n,m}^{tar,w}$ and $\tau _{s,l,n,m}^{e,tar}$) of the paths in (\ref{eqn_3b}) exhibit a two-stage pattern. The random phase ($\psi_{s,l,n,m}^{i,j}$) and Doppler shift (${{f_{d,s,l,n,m}}}$) in (\ref{eqn_3b}) represent the joint effect of two sub-channels. As for targets with known coordinates ($[x_l,y_l,z_l]$), the 3D distances of TX$_s$-target$_l$ or target$_l$-RX$_s$ sub-channels are calculated as
\begin{equation}
\label{equ_6}
{d_{{e,tar/tar,w}}^{\rm{3D}}}= {\left\| {({x_{e/w}},{y_{e/w}},{z_{e/w}})- {x_l},{y_l},{z_l}} \right\|_2},
\end{equation}
where $[{x_{e/w}},{y_{e/w}},{z_{e/w}}]$ denotes the coordinate of TX/RX antenna $e/w$. $\left\| \cdot \right\|_2$ represents the Euclidean norm. The LoS ZAoA, AAoA, ZAoD ($\theta _{s,l,{\rm{los}}}^{\rm{zod}}$), and AAoD ($\phi _{s,l,{\rm{los}}}^{\rm{aod}}$) can be calculated under the global coordinate system. When the LoS clusters propagating through the targets are considered in sensing channel, these geometric LoS angles can be employed directly as the cluster angles. In this context, the LoS delay of cascade sensing channel are calculated by
\begin{equation}
\label{equ_7}
{\tau _{{s,l,\rm{los}}}} ={\tau _{s,l,{\rm{los}}}^{e,tar}} + {\tau _{s,l,{\rm{los}}}^{tar,w}}
=({d_{{e,tar}}^{\rm{3D}}} + {d_{{tar,w}}^{\rm{3D}}})/c,
\end{equation}
where $c$ denotes the speed of electromagnetic waves.

Given that the targets are known, the sensing parameters generated by these targets can be feedback to the communication channel to enhance modeling accuracy. The $l^{th}$ target, as a potential communication scatterer, contributes to the communication propagation paths as ${H_{c,q,l,p}}(t,\tau_c)$. Case 1: when communication TX is integrated with mono-static sensing, ${H_{c,q,l,p}}(t,\tau_c)$ can reuse the path departure angles of ${H_{s,w,l,e}}(t,\tau_s)$. Case 2: when communication RX is integrated, the path arrival angles can reuse. Case 3: when communication TX/RX are integrated with bi-static sensing, propagation parameters of ${H_{c,q,l,p}}(t,\tau_c)$ and ${H_{s,w,l,e}}(t,\tau_s)$ can be considered equivalent. Note that the model expression and other propagation parameters in ${H_{c,q,l,p}}(t,\tau_c)$ still adhere to 3GPP standards, as denoted in (\ref{eqn_3a}). 

To model the sharing feature of ISAC channel, this letter proposes a novel parameter $\mathop {\bf{S}}$ with dimensions $1\times L$ for the shared cluster extraction from these targets. The calculations based on (\ref{eqn_1}) and (\ref{eqn_2}) are expressed as
\begin{equation}
\label{eqn_8}
H_{c,q,p}^{{\rm{sc}}}(t,\tau_c)=\mathop {\bf{S}}\limits_{1 \times L} \cdot \mathop {{\bf{H}}_{c,q,p}^{{\rm{tar}}}}\limits_{L \times 1}(t,\tau_c),
\end{equation}
\begin{equation}
\label{eqn_9}
H_{s,w,e}^{{\rm{sc}}}(t,{\tau _s})=\mathop {\bf{S}}\limits_{1 \times L}  \cdot \mathop {{\bf{H}}_{s,w,p}^{{\rm{tar}}}}\limits_{L \times 1} (t,{\tau _s}),
\end{equation}
\begin{equation}
\label{eqn_10}
H_{s,w,p}^{{\rm{nsc}}}(t,{\tau _s})=\mathop {({\bf{I}} - {\bf{S}})}\limits_{1 \times L}  \cdot \mathop {{\bf{H}}_{s,w,e}^{{\rm{tar}}}}\limits_{L \times 1} (t,{\tau _s}),
\end{equation} 
where $\mathop {{\bf{H}}_{c,q,p}^{{\rm{tar}}}(t,\tau_c)}=[H_{c,q,l,p}(t,\tau_c)]_{L\times 1}$ and $\mathop {{\bf{H}}_{s,w,e}^{{\rm{tar}}}(t,\tau_s)}=[H_{s,w,l,e}(t,\tau_s)]_{L\times 1}$. ${\bf{I}}=\left [ 1,1,\cdots,1,1 \right ]_{L\times 1}$ denote the unit vector. $\mathop {\bf{S}}$ contains $L$ nonnegative real-values, i.e.,
\begin{equation}
\label{eqn_11}
\mathop {\bf{S}}=\left [ s_1,\cdots,s_l,\cdots,s_L \right ], s_l=0\ \rm{or}\ 1.
\end{equation}
$s_l=1$ means that the $l^{th}$ target acts in both communication and sensing channels, ${H_{s,w,l,e}}(t,\tau_s)$ belongs to $H_{s,w,p}^{{\rm{sc}}}(t,{\tau _s})$ and communication clusters/paths contributed by this target are generated. On the contrary, $s_l=0$ represents the $l^{th}$ target only serves as sensing scatterer, resulting in ${H_{s,w,l,e}}(t,\tau_s)$ belonging to $H_{s,w,p}^{{\rm{nsc}}}(t,{\tau _s})$. 

In (\ref{eqn_3}), the relationship of shared cluster number can be expressed as $N_{c0}=L_{s0}\times N_{s0}$ ($L_{s0}=\sum \bf{S}$).
In sensing channel, the LoS cluster has the highest power and provides more effective position information. $N_{c0}=L_{s0}$ when the sensing channel only characterizes cascaded LoS clusters ($N_{s0}=1$). 
Moreover, $H_{c,q,p}^{{\rm{nsc}}}(t,{\tau _c})$ in (\ref{eqn_1}) is stochastically modeled according to the traditional way.





\section{Modeling Implementation and numerical analysis}\label{section3}

\subsection{ISAC Modeling Implementation}

\begin{figure}[!h]
\centering
\includegraphics[width=3.4in]{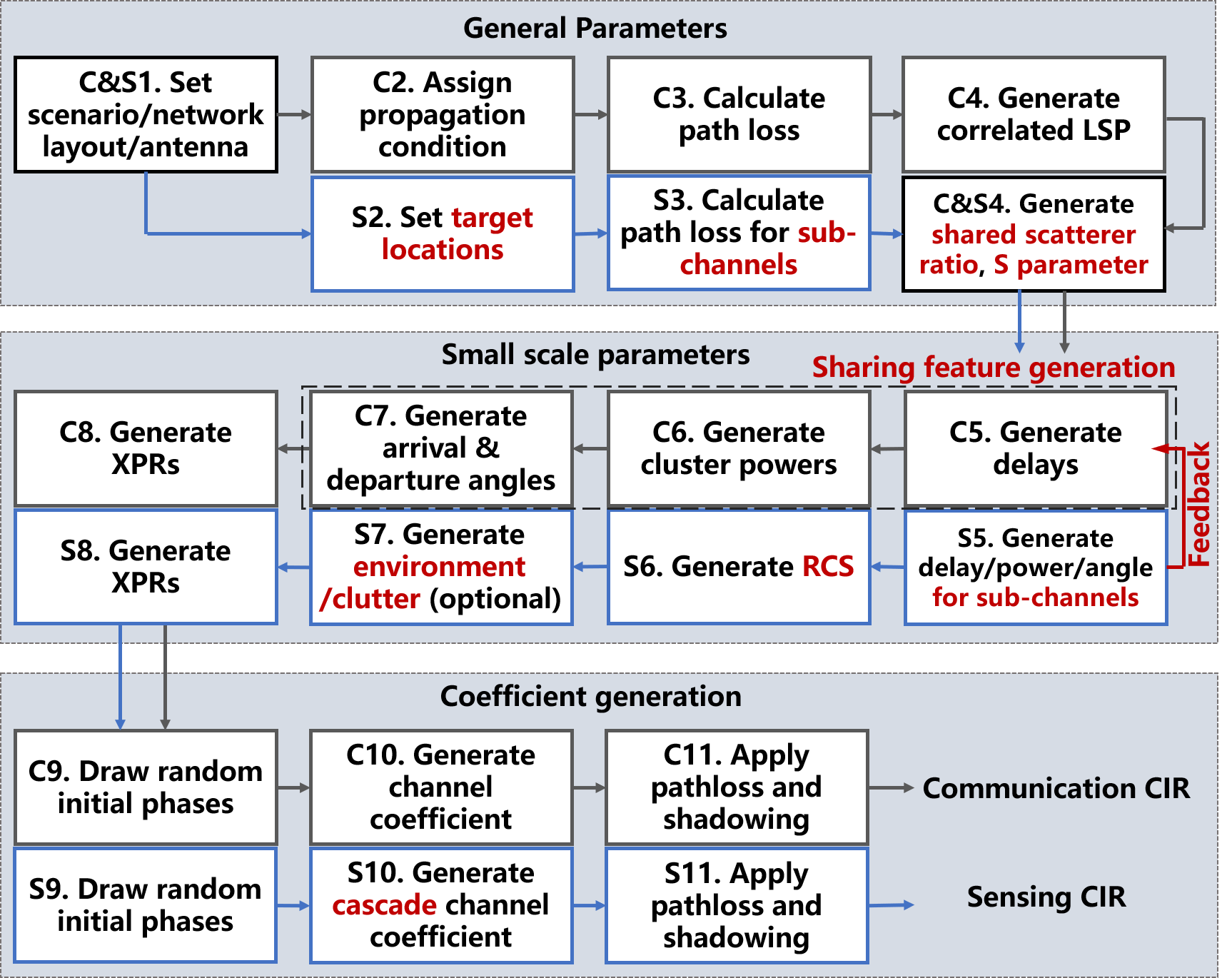}
\caption{ISAC channel model implementation framework.}
\label{fig_2}
\end{figure}

Based on the proposed ISAC channel model, this letter extends the 3D GBSM and proposes an ISAC channel implementation framework, as illustrated in Fig. \ref{fig_2}. This framework, built upon the traditional communication modeling flow in accordance with 3GPP \cite{3gpp38901}, incorporates the generation of sensing channels and the modeling of shared feature. Parameters generation steps that deviate from standard specifications are highlighted in red in the diagram. 


\emph {1) Sensing channel realization:} As illustrated by the blue flowcharts in Fig. \ref{fig_2}, the position ($[x_l,y_l,z_l]$) and velocity ($\overline{v}$) of sensing targets are preset in \emph{step S2}. Then, the distances and LoS angles of sub-channels are calculated. The path loss in \emph{step S3} can be generated as the multiplication of two independent sub-channels according to \cite{3gpp38901} or calculated based on free space radar equation. In \emph{step S5}, traditional channel methodologies apply to both sub-channels, with cluster delays, powers, and angles generated. The RCS coefficient in \emph{step S6} can be initially implemented as fixed/stochastic values in the preset range depending on the typical object at a specific frequency. The generation of environment/clutter serves as an optional module in \emph{step S7}, which this letter doesn't discuss in detail. \emph{Step S10} generates cascade channel coefficient based on (\ref{eqn_4}). The other sensing parameters are realized in the stochastic way according to the traditional 3D GBSM.


\emph {2) Communication channel realization:} As illustrated by the gray flowcharts in Fig. \ref{fig_2}, the primary parameters of the communication channel, such as large-scale delay spread (DS), angle spread (AS), shadow fading (SF), Ricean K-factor, etc., are generated in accordance with the 3GPP standards. Additionally, to model the sharing feature in the ISAC channel, the expected shared scatterer ratio ($\sum {\bf{S}}/L$) is set in \emph{step C\&S4}. Focusing on the cascade LoS clusters in sensing channel (i.e., $N_{s0}=1$), this letter provides a method for selecting $N_{c0}$ shared cluster pairs and determining ${\bf{S}}$ by comparing specific parameters (such as AAoD and ZAoD) of sensing clusters (${{H_{s,w,l,e}}}$) generated in \emph{step S5} and stochastic communication clusters ($H_{c,q,p}^{{\rm{nsc}}}$) generated in \textit{steps C5-C7}. Taking integrated case 1 defined in Section \ref{section2}-B as an example, the selection criteria are shown as formulas (\ref{equ_12})-(\ref{equ_14}).
\begin{gather}
\label{equ_12}
{\rm{ga}}{{\rm{p}}^{{\rm{aod}}}}{\rm{(}}{l},{n}{\rm{) = }}\left| {\phi _{s,l,{\rm{los}}}^{\rm{aod}} - \phi _{c,n}^{{\rm{aod}}}} \right| \in [0,2\pi ],\\
\label{equ_13}
{\rm{ga}}{{\rm{p}}^{{\rm{zod}}}}{\rm{(}}{l},{n}{\rm{) = }}\left| {\theta _{s,l,{\rm{los}}}^{\rm{zod}} - \theta _{c,n}^{{\rm{zod}}}} \right| \in [0,\pi ],\\
\label{equ_14}
{({l},{n})^{{n_{c0}}}}{\rm{ = }}\mathop {{\rm{argmin}}}\limits_{{l},{n}} \left\{ {\frac{1}{2}\left( {\frac{{{\rm{ga}}{{\rm{p}}^{{\rm{aod}}}}}}{{2\pi }} + \frac{{{\rm{ga}}{{\rm{p}}^{{\rm{zod}}}}}}{\pi }} \right)} \right\},
\end{gather}
where ${\rm{ga}}{{\rm{p}}^{{\rm{aod}}}}$ and ${\rm{ga}}{{\rm{p}}^{{\rm{zod}}}}$ denote the AAoD and ZAoD gap between $l^{th}$ sensing target and $n^{th}$ communication stochastic cluster. $\phi _{c,n}^{{\rm{aod}}}$ and $\theta _{c,n}^{{\rm{zod}}}$ represent cluster AAoD and ZAoD of $n^{th}$ communication cluster. $({l},{n})^{{n_{c0}}}$ represent ${n_{c0}}^{th}$ shared cluster pair. According to (\ref{equ_14}), the $N_{c0}$ pairs of shared cluster labels $(l,n)$ are selected sequentially without repetition, and $s_l=1$ in (\ref{eqn_11}) are set accordingly.

To balance the statistical nature of the communication channel and the deterministic features of the sensing channel, the AAoD and ZAoD of sensing clusters in $({l},{n})^{{n_{c0}}}$ feedback and replace corresponding parameters of communication clusters, i.e., $\phi _{c,n}^{{\rm{aod}}}=\phi _{s,l,{\rm{los}}}^{\rm{aod}}$ and $\theta _{c,n}^{{\rm{zod}}}=\theta _{s,l,{\rm{los}}}^{\rm{zod}}$. The remaining parameters of shared communication clusters ($H_{c,q,p}^{{\rm{sc}}}$) are generated in reverse according to 3GPP standards and combined with $H_{c,q,p}^{{\rm{nsc}}}$. Finally, the communication channel coefficients are generated according to the flowchart. Note that when the shared parameter in \emph{step C\&S4} is set to 0, the communication and sensing channels will be independent. This configuration is compatible with traditional communication channel implementations.

\subsection{Numerical Analysis}

\begin{figure}[!h]
\centering
\includegraphics[width=3.2in]{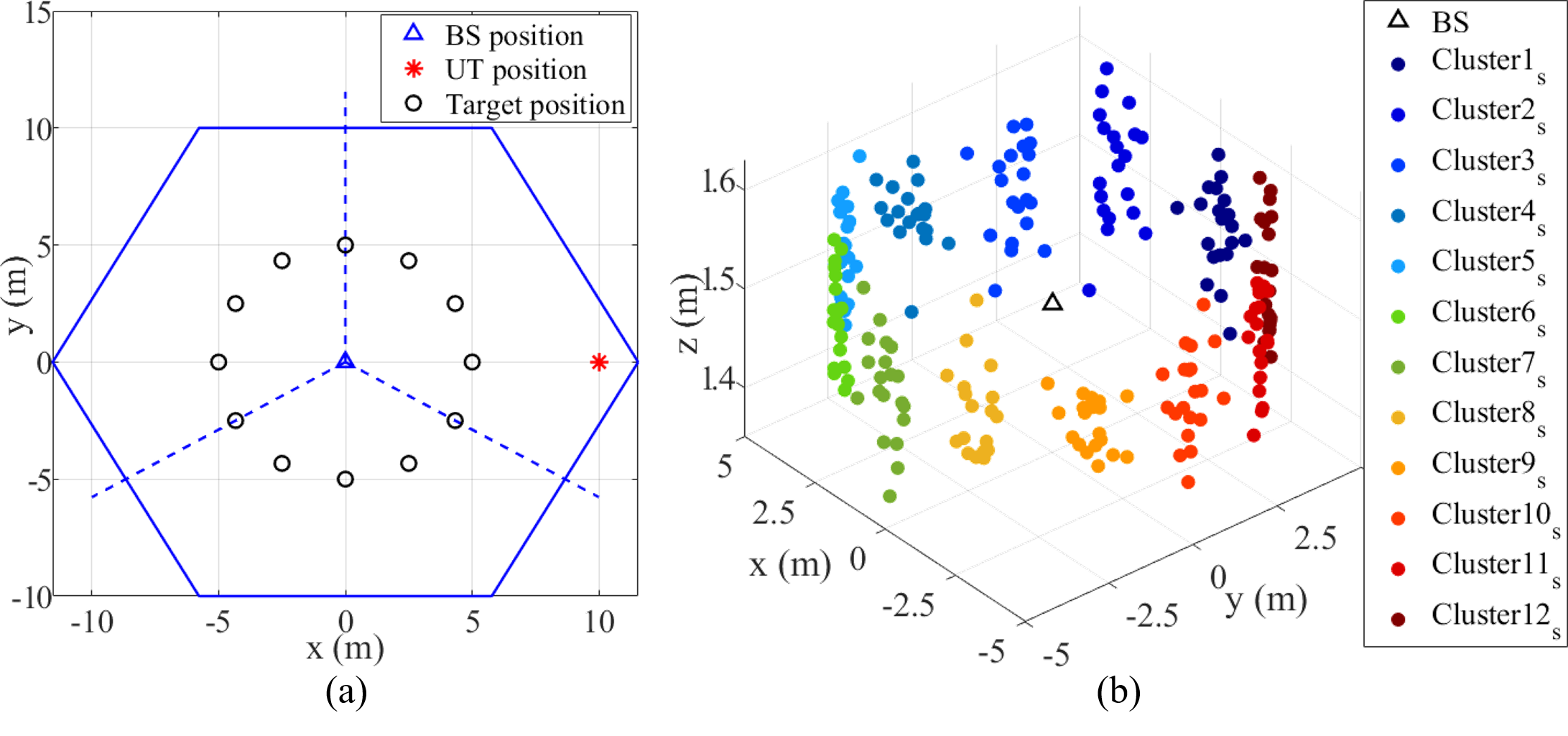}
\caption{Simulation positions of (a) TX, RX, and targets and (b) sensing paths.}
\label{fig_3}
\end{figure}

This letter performs simulations to valide the proposed modeling framework for ISAC channel. In \emph{step C\&S1} of simulations (as shown in Fig. \ref{fig_2}), the coordinate of the BS (sensing TX/RX and communication TX) is set as $(0,0,1.5)\ \text{m}$, and the UT (communication RX) is set as $(10,0,1.5)\ \text{m}$. The Indoor Hotspot (InH) scenario is configured as the baseline scenario. In \emph{step S2}, we set up 12 targets ($L=12$) around the origin. These targets are $5\ \text{m}$ from the origin, $1.5\ \text{m}$ in height and spaced with $30^\circ$ horizontally. Only LoS clusters are set for each target, i.e., $N_{s0}=1$. Fig. \ref{fig_3}(a) shows the 2D simulation positions, where the blue lines denote the area of communication sectors. Fig. \ref{fig_3}(b) presents the 3D positions of mono-static sensing paths, each belonging to one of the 12 target clusters. The simulation results demonstrate that the sensing clusters and paths obtained using the proposed modeling framework accurately correspond to the target positions.


To reflect the parameter relationship between communication and sensing shared clusters, we set shared scatterer ratio to $1/6$ in \emph{step C\&S4} and the shared clusters amount to 2. As an example, we consider the communication LoS scenario in one of our simulations. To ensure consistency with the 3GPP guidelines \cite{3gpp38901}, we remove clusters with power levels more than 25 dB below the maximum cluster power. This process results in 13 effective communication clusters in one simulation. The simulation clusters/paths of sensing and communication channels in the AAoD and ZAoD dimensions are shown in Fig. \ref{fig_3}. In this simulation result, sensing cluster 1 and communication LoS cluster 1 are shared, as well as sensing cluster 11 and communication NLoS cluster 4. These shared clusters have the same cluster centroid departure angles.

\begin{figure}[!h]
\centering
\includegraphics[width=3.4in]{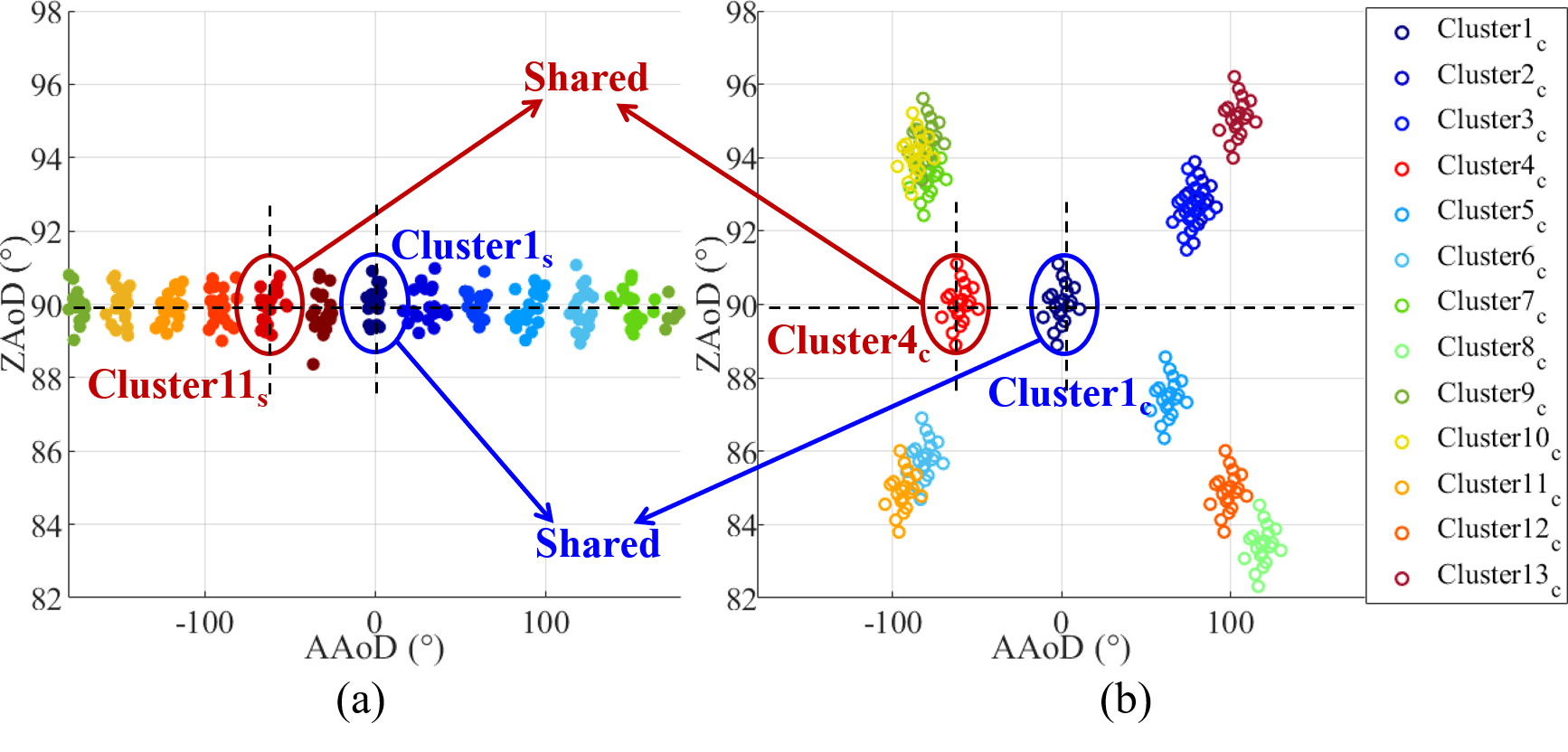}
\caption{Simulation clusters/paths of (a) sensing and (b) communication channels in dimensions of AAoD($^\circ$) and ZAoD($^\circ$).}
\label{fig_4}
\end{figure}

\begin{figure}[h]
\centering
\subfloat[]{\includegraphics[width=1.75in]{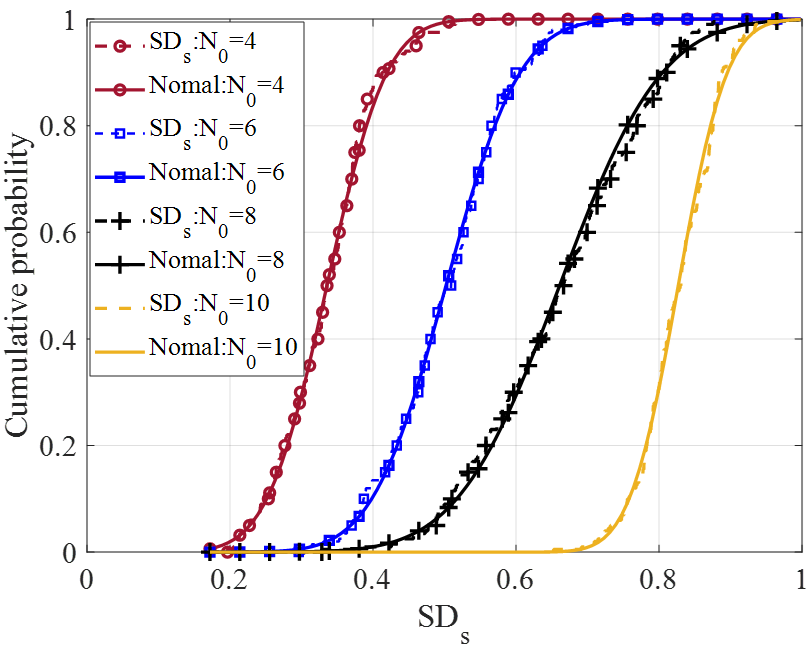}\label{fig_5a}}
\subfloat[]{\includegraphics[width=1.75in]{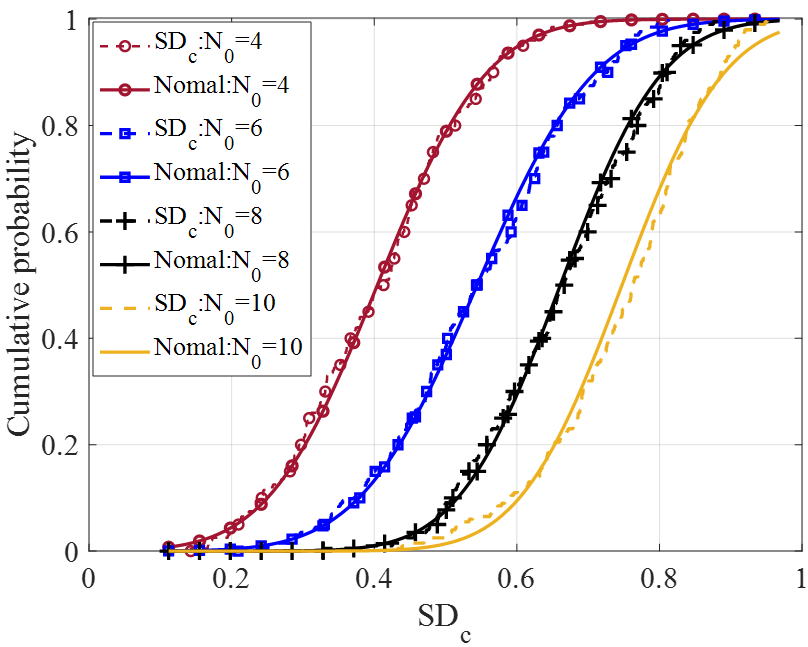}\label{fig_5b}}
\caption{Simulation CDFs of SD in (a) sensing and (b) communication channels.}
\label{fig_5}
\end{figure}

To further capture shared statistical characteristics, we simulate the shared degree (SD) metric, which has been defined in \cite{liu2022shared}. The SD represents the ratio of shared cluster power to the total cluster power. While keeping remaining simulation parameters constant, we conduct 200 simulations with varying shared cluster numbers of 4, 6, 8, and 10, respectively. The statistical Cumulative Distribution Functions (CDFs) of sensing and communication SD, respectively denoted as ${\rm{SD}}_s$ and ${\rm{SD}}_c$, are illustrated in Fig. \ref{eqn_5}. It is evident that an increasing shared cluster number leads to higher SD values. Since the distances between the preset targets and the BS are uniformly consistent in this letter, the LoS cluster power for each target is also similar. Therefore, Under identical shared cluster settings, the sensing SD values exhibit greater concentration, which is reflected in steeper CDF curves in Fig. \ref{fig_5a}.
 
\section{Conclusion}\label{section4}
This letter extends the 3D GBSM and proposes an ISAC channel model that incorporates the sharing feature. This model smoothly aligns with the 3GPP standardization approach. Different from the existing ISAC model, communication and sensing channels are modeled as the shared and non-shared components superimposed in the proposed channel model. For targets of interest, the sensing channel is depicted as a cascade of TX-target, RCS, and target-RX, with RCS modeled as a 2×2 array with different polarization directions. To capture the sharing feature of ISAC channel, this letter introduces a novel parameter $\bf{S}$ for the shared targets/clusters extraction. Based on the proposed model, we present an ISAC channel implementation framework that facilitates the joint generation of communication and sensing channels. To validate the model performance, this letter conducts simulations to examine the positions of sensing channel clusters/paths, the relationship between shared parameters such as cluster angles, and the statistical SD. The results demonstrate that the proposed channel model effectively portrays the practical sharing feature and can support the design of ISAC systems.



\bibliography{reference}




 

\vfill

\end{document}